\documentclass[article, shortnames, nojss]{jss}

\makeatletter
\def\maxwidth{ %
  \ifdim\Gin@nat@width>\linewidth
    \linewidth
  \else
    \Gin@nat@width
  \fi
}
\makeatother

\usepackage[]{graphicx}\usepackage[]{color}
\usepackage{Sweave}
\usepackage{amsmath}
\usepackage{amssymb}
\usepackage{algorithm}
\usepackage{algorithmic}
\usepackage{tabulary}
\usepackage[usenames,dvipsnames]{xcolor}
\usepackage{hyperref}
\usepackage{bm}
\usepackage{booktabs}
\usepackage{threeparttable}
\usepackage{microtype}

\newcommand\Tstrut{\rule{0pt}{2.6ex}}         % = `top' strut
\newcommand\Bstrut{\rule[-0.9ex]{0pt}{0pt}}   % = `bottom' strut

\author{Vasiliki Koutra\\ King's College \\London \And 
Olga Egorova \\ King's College \\London \And 
Steven G. Gilmour\\ King's College \\London\And 
Luzia A. Trinca \\ S\~{a}o Paulo \\State University}

\title{\pkg{MOODE}: An \proglang{R} Package for Multi-Objective Optimal Design of Experiments}

\Plainauthor{Vasiliki Koutra, Olga Egorova, Steven G. Gilmour, Luzia A. Trinca}
\Plaintitle{MOODE: An R Package for Multi-Objective Optimal Design of Experiments}
\Shorttitle{\pkg{MOODE}: Multi-Objective Optimal Design of Experiments}

\Abstract{
We describe the \proglang{R} package \pkg{MOODE} and demonstrate its use to find multi-objective optimal experimental designs. Multi-Objective Optimal Design of Experiments (MOODE) targets the experimental objectives directly, ensuring that the full set of research questions is answered as economically as possible. In particular, individual criteria aimed at optimizing inference are combined with lack-of-fit and MSE-based components in compound optimality criteria to target multiple and competing objectives reflecting the priorities and aims of the experimentation.  The package implements either a point exchange or coordinate exchange algorithm as appropriate to find nearly optimal designs. We demonstrate the functionality of \pkg{MOODE} through the application of the methodology to two case studies of varying complexity.}    


\Address{
 Vasiliki Koutra, Olga Egorova, Steven G. Gilmour\\
  Department of Mathematics\\
  King's College London\\
  Strand, London, WC2R 2LS, UK\\
  E-mail: \email{\{Vasiliki.Koutra,O.Egorova, Steven.Gilmour\}@kcl.ac.uk}\\[1ex]
Luzia A. Trinca\\
  Departamento de Biodiversidade e Bioestat{\'i}stica \\
  Unesp - S{\~a}o Paulo State University \\
  Botucatu, CEP 18618-689 \\
  Brazil \\
  E-mail: \email{luzia.trinca@unesp.br}
}

\IfFileExists{upquote.sty}{\usepackage{upquote}}{}

\begin{document}
\section[Introduction]{Introduction}
\label{sec:intro}

A well-designed and executed experiment is an efficient way of understanding the impacts of different interventions on a process \citep{BHH2005}, allowing interpretable conclusions to be made regarding the shape and strength of the relationships between controllable treatment factors and the measured response. Response Surface Methodology \citep{BoxandWilson1951} aims to build models to describe these relationships, most commonly second-order polynomials, with the usual ultimate aim of finding an optimal response. Planning good experiments to estimate such models and achieve the experimental goals requires specification of the anticipated response model prior to experimentation. As there is usually uncertainty about the correct polynomial form to assume, this is a well-recognised drawback of model-based optimal design \citep[][ch.~13]{Box1987empirical}. In addition, there may often be a tension between the size of the assumed model (e.g., degree of polynomial, inclusion of interactions) and the size of the experiment, with larger models requiring experiments with a greater number of runs to provide useful estimation and inference.

Hence, an ideal experiment would address these concerns, perhaps via precise estimation of the parameters in an assumed model thought most likely a priori, whilst being able to detect lack of fit from this model and reduce bias resulting from excluded polynomial terms. Recently, \citet{EgorovaGilmour2022} proposed a class of compound criteria for exactly this purpose, reflecting the uncertainty in the assumed model and the need to tension between estimation of this model and detection of, and protection against, alternatives. The \pkg{MOODE} package (Multi-Objective Optimal Design of Experiments) finds designs within this framework. It is also desirable for the design to allow estimation of background error from replication, unbiased by the choice of statistical model, to allow formal statistical inference. The \pkg{MOODE} package therefore incorporates criteria focused on statistical inference \citep{GilmourandTrinca2012}. The package is available to download from \url{https://github.com/vkstats/MOODE}.

\subsection{Response surface models}\label{preliminaries}
An $n$-run experiment is to be performed when it is assumed that the response surface is smooth enough to be described via a polynomial regression model in $k$ quantitative factors. We follow the common convention in experimental design of referring to all controllable features of the experiment as factors, including continuous variables, rather than reserving this terminology for discrete categorical variables. Further, it is assumed that each factor will be set to values from a discrete set of levels $\mathcal{L}_k\subset \mathcal{R}$. Such a restriction is common both in practical experiments, where only a small number of values of the factor may be achievable, and when finding optimal experimental designs even for continuous factors (where for linear models careful choice of the set of factor levels can ensure no loss of efficiency from this assumption). Each distinct combination of factor values applied in the experiment is denoted as a treatment, with $t$ denoting the number of treatments used.  

To set-up the parameters for a new experiment, the \code{mood} function in the \pkg{MOODE} package can be used, e.g., as below for $k=2$ factors, each at three levels, with $n=25$ runs.
\begin{CodeInput}
library(MOODE)
mood_ex <- mood(K = 2, Levels = 3, Nruns = 24)
\end{CodeInput}
Further inputs to the \code{mood} function will be introduced in this and the following sections.

The package assumes observation $Y_i$ from the $i$th run ($i = 1, \ldots, n$) is modelled as
\begin{equation}\label{eq:primary_model}
\begin{split}
    Y_i & = \beta_0 + \sum_{j = 1} ^ {p} \beta_j x_{ij} + \varepsilon_{i} \\
    & = \beta_0 + \boldsymbol{x}_{i1}^\top\boldsymbol{\beta}_1 + \varepsilon_{i}\,,
\end{split}
\end{equation}
with $\beta_0$ and $\boldsymbol{\beta}_1^\top = (\beta_1, \ldots, \beta_{p})$ containing unknown parameters to be estimated, and $\boldsymbol{x}_{i1}^\top = (x_{i1}, \ldots, x_{ip})$ -- the values of the $p$ predictors for the $i$th run. The unit effects (errors) $\varepsilon_{i}$ have expectation zero and constant variance $\sigma^2$, with $\varepsilon_{i}, \varepsilon_{i^\prime}$ assumed independent for $i \neq i^\prime$. The $p$ predictors may include linear terms in the $k$ factors, higher-order polynomial terms and interactions.

Two further models are also incorporated into the design selection via the \pkg{MOODE} package. \citet{Box1959} introduced the idea of incorporation of discrepancy between the assumed response surface model and an encompassing ``true'' model. The encompassing model contained addition polynomial terms; here we assume there are $q$ such terms, typically higher-order monomials, possibly including interactions, which we will label $x_{i(p+1)},\ldots, x_{i(p+q)}$. The encompassing model can therefore be written as
\begin{equation}\label{eq:primary_potential_model}
\begin{split}
    Y_i & = \beta_0 + \sum_{j = 1} ^ {p} \beta_j x_{ij} + \sum_{j = p+1} ^ {p+q} \beta_j x_{ij} +\varepsilon_i \\
    & = \beta_0 + \boldsymbol{x}_{i1}^\top\boldsymbol{\beta}_1 + \boldsymbol{x}_{i2}^\top\boldsymbol{\beta}_2 + \varepsilon_i\,,
\end{split}
\end{equation}

where $\boldsymbol{x}_{i2}^\top = (x_{i(p+1)}, \ldots, x_{i(p+q)})$ holds the additional $q$ polynomial terms, with associated parameters $\boldsymbol{\beta}_2^\top = (\beta_{p+1}, \ldots, \beta_{p+q})$. Model~\eqref{eq:primary_model} is clearly a special case of model~\eqref{eq:primary_potential_model} with $\boldsymbol{\beta}_2 = \boldsymbol{0}_q$. \citet{DuMouchel1994} labelled the polynomial terms in the assumed model as \textbf{primary} and the additional terms in the encompassing model as \textbf{potential}.

Specification of the assumed primary model and possible additional potential terms in the \pkg{MOODE} package is via the optional argument \code{model_terms}, a list with named elements. There are two methods for specifying the primary and potential models. 
\begin{enumerate}
\item List entries \code{primary.model} and \code{potential.model} may be provided, specifying the primary and potential models via strings.
\item Alternatively, list entries \code{primary.terms} and \code{potential.terms} may be provided to directly specify model terms. 
\end{enumerate}
For example, both the following two approaches specify the same primary model~\eqref{eq:primary_model} with linear terms ($p = 3$), with additional potential terms in~\eqref{eq:primary_potential_model} being quadratic effects ($q = 2$) for $k=2$ factors.
\begin{CodeInput}
    ex.mood <- mood(K = 2, Levels = 3, Nruns = 24, model.terms = list(
    primary.model = "main_effects", potential.model = "quadratic_terms"))
    
    ex.mood <- mood(K = 2, Levels = 3, Nruns = 24, model.terms = list(
    primary.terms = c("x1", "x2"), potential.terms = c("x12", "x22")))
\end{CodeInput}
If both \code{*.model} and \code{*.terms} are specified, the model specified via \code{*.terms} takes precedence. An intercept is always included in the primary model. The default primary model contains the linear terms in each factor (\code{primary.model = "main_effects"}, $p = k$), with no potential terms by default (\code{potential.terms = NULL}, $q = 0$).

The final model considered in the \pkg{MOODE} package is the full treatment model
\begin{equation}\label{eq:full_treat_model}
Y_i = \mu_{m(i)} + \varepsilon_{ti}\,,
\end{equation}
where $m(i) \in \{1,\ldots,t\}$ denotes the treatment applied to the $i$th unit. No simplifying relationship is assumed in model~\eqref{eq:full_treat_model} for the mean responses from different treatments, and there are $t$ different model parameters that require estimation $\boldsymbol{\mu}^\top = (\mu_1, \ldots, \mu_t)$. Hence the $\varepsilon_{ti}$ are the unit-to-unit differences in the experiment. Clearly, the mean functions in models~\eqref{eq:primary_model} and~\eqref{eq:primary_potential_model} from assumptions
$$
\mu_{m(i)} = \beta_0 + \boldsymbol{x}_{i1}^\top\boldsymbol{\beta}_1\,
$$
and
$$
\mu_{m(i)} = \beta_0 + \boldsymbol{x}_{i1}^\top\boldsymbol{\beta}_1 + \boldsymbol{x}_{i2}^\top\boldsymbol{\beta}_2\,,
$$
respectively. The error terms $\varepsilon_i$ in these two models measure both unit-to-unit variation and model lack-of-fit.

For a design including replication ($n>t$),  \citet{GilmourandTrinca2012} discussed the advantages of basing inference on the estimate of $\sigma^2$ obtained from using the residual mean square from model~\eqref{eq:full_treat_model}. The correctness of this estimate depends only on the minimal assumption of additive treatment and unit effects and not on which function is used to approximate the relationship of interest. The same estimate can be obtained from fitting model~\eqref{eq:primary_model} and decomposing the residual sum of squares, with $n-p$ degrees of freedom into components for `lack of fit' (with $t-p$ degrees of freedom) and `pure error' (with $n-t$ degrees of freedom). The latter quantity results from the replication in the experiment.

Each of models~\eqref{eq:primary_model},~\eqref{eq:primary_potential_model} and~\eqref{eq:full_treat_model} can be written in matrix form, collecting together the responses from the $n$ experimental units, as
\begin{align}
    \boldsymbol{Y} & = \boldsymbol{1}_n\beta_0 + X_1\boldsymbol{\beta}_1 + \boldsymbol{\varepsilon} \nonumber\\ 
    & = X\boldsymbol{\beta} + \boldsymbol{\varepsilon}\,,\label{eq:primary_matrix} \\
    \boldsymbol{Y} & = \boldsymbol{1}_n\beta_0 + X_1\boldsymbol{\beta}_1 + X_2\boldsymbol{\beta}_2 + \boldsymbol{\varepsilon} \label{eq:primary_potential_matrix}\\
\boldsymbol{Y} & = X_f\boldsymbol{\mu} + \boldsymbol{\varepsilon}_t \,,\label{eq:treat_matrix} 
\end{align}
with $\boldsymbol{Y}^\top = (Y_1, \ldots, Y_n)$, $X = [\boldsymbol{1}_n\,~X_1]$, $\boldsymbol{\beta} = (\beta_0, \boldsymbol{\beta}_1^\top)^\top$, $\boldsymbol{\varepsilon}^\top = (\varepsilon_1, \ldots, \varepsilon_n)$, $\boldsymbol{\varepsilon}_t^\top = (\varepsilon_{t1}, \ldots, \varepsilon_{tn})$, $\boldsymbol{1}_n$ being the $n$-vector of all-ones, model matrices $X_1$ and $X_2$ having $i$th row $\boldsymbol{x}_{i1}^\top$ and $\boldsymbol{x}_{i2}^\top$, respectively, and $X_f$ being the model matrix for the full treatment model with $(i,j)$th entry being equal to 1 if the $j$th treatment is applied to the $i$th unit and being equal to 0 otherwise.

\subsection[Related R packages]{Related \proglang{R} packages}

The current \proglang{R} ecosystem for design of experiments is overviewed by the CRAN Task View (\url{https://cran.r-project.org/web/views/ExperimentalDesign.html}). The main packages tailored to construction of optimal designs for response-surface models (amongst other aims) are \pkg{AlgDesign} \citep{AlgDesign}, \pkg{skpt} \citep{skpr} and \pkg{OptimalDesign} \citep{OptimalDesign}. All three packages find good subsets of points from a \textit{candidate list} of possible points for inclusion in the design (typically constructed as a full factorial in the possible levels of all the factors).

The \pkg{AlgDesign} package is the most widely used of these (as measured by CRAN downloads), and implements point exchange algorithms for the construction of optimal designs for linear models under a variety of precision-based criteria (eg., $A$-optimality and $D$-optimality). Therefore, the designs generated are focussed on point estimation or prediction from a single assumed response model.  

Package \pkg{skpr} also uses a point exchange algorithm to generate designs under precision-based criteria. In addition, the package can find designs under ``Alias-optimality'' \citep{JonesNachtsheim2011b}, minimising the weighted cross-products between $X_p$ and $X_q$ in~\eqref{eq:primary_potential_matrix}; see also Section~\ref{sec:model-robust}. Designs can also be generated under user-defined criteria, which are functions of the model matrix $X_p$.

$A$- and $D$-optimal designs, among others, can also be generated for linear models using the \pkg{OptimalDesign} package. This package implements a variety of different algorithmic approaches, including point exchange, mixed-integer programming and quadratic approximations. A particular focus is the incorporation of constraints on the design region.

A few packages also address model uncertainty in design choice, including ``optimum-on-average'' criteria for nonlinear models (\pkg{ICAOD}; \citealp{ICAOD}); $T$-optimality for model-discriminating designs (\pkg{rodd}; \citealp{rodd}); and general Bayesian criteria using approximate coordinate exchange (\pkg{acebayes}; \citealp{acebayes}) and follow-up designs for model discrimination (\pkg{OBsMD}; \citealp{OBsMD}). No packages address the same range of inference aims and model uncertainty as \pkg{MOODE}.   

\subsection{Outline}

The focus of this paper is the introduction of the \pkg{MOODE} package and its demonstration on a number of applications for response surface experiments with a relatively small number of runs. In Section~\ref{sec:criteria}, the main design selection criteria are introduced and their combination through a compound objective function is described. The implemented algorithms for design optimisation, point and coordinate exchange, are described in Section~\ref{sec:algorithms}. In Section~\ref{sec:examples} the use of the \pkg{MOODE} package to find optimal designs is demonstrated on two examples, where compromises across the competing objectives and criteria are illustrated. We conclude in Section~\ref{sec:disc} with a short discussion and some recommendations. Functionality of the \pkg{MOODE} package is described and demonstrated throughout.

\section[sec1]{Design optimality criteria}
\label{sec:criteria}

The \pkg{MOODE} package implements design optimality criteria that can be categorised as emphasising model estimation and inference, model sensitivity, and model robustness. The implemented criteria in each of these classes are described below in Sections~\ref{sec:model-inf}-~\ref{sec:model-robust} and summarised in Table~\ref{tab:criteria} with the associated \pkg{MOODE} functions. See \citet{EgorovaGilmour2022} for more methodological details. 

The main purpose of the \pkg{MOODE} package is to find designs under compound criteria formed as weighted products of objective functions composed of these individual criteria, as explained in Section~\ref{sec:compound}. Throughout, we assume a design is given by a $n\times k$ matrix $\mathcal{D}$ with $(i,j)$-th entry $x_{ij}$ where we assume that the first $k$ predictors correspond to the continuous factors being controlled in the experiment.

Each criterion is implemented via the argument \code{criterion.choice} to the \code{mood} function, and is also available through a corresponding \code{criteria.*} function that takes three main arguments: primary and potential model matrices $X_1$ and $X_2$ for a given design and an object of class \code{mood} specifying the experiment arguments, including the primary and potential models. 

\begin{table}
 \begin{threeparttable}
\begin{center}
\begin{tabular}{lll} 
\hline
Criterion & Objective function & \pkg{MOODE} \\ \hline
\multicolumn{3}{l}{\textbf{Estimation and inference}} \\
$D_S$-optimality & $\phi_{D_S}(\mathcal{D}) = |[X_1^\top(I_n - \frac{1}{n}J_n)X_1]^{-1}|$ & \code{criteria.GD} \\
$L$-optimality & $\phi_L(\mathcal{D})  =  \mbox{tr}\{L^\top[X_1^\top(I_n - \frac{1}{n}J_n)X_1]^{-1}L\}$ & \code{criteria.GL} \\
$(DP)_S$-optimality & $\phi_{(DP)_S}(\mathcal{D}) =  F_{p,d;1-\alpha}^p\phi_D(\mathcal{D})$ & \code{criteria.GDP} \\
$LP$-optimality & $\phi_{LP}(\mathcal{D}) =  F_{1,d;1-\alpha}\phi_L(\mathcal{D})$ & \code{criteria.GLP} \\
\textbf{Sensitivity} \\
LoF-$DP$-optimality\tnote{1} & $\phi_{LoF-DP}(\mathcal{D}) = F^q_{q, d; 1-\alpha_{L}} |L + \frac{1}{\tau^2}I_q|^{-1}$ & \\
LoF-$LP$-optimality\tnote{2} & $\phi_{LoF-LP}(\mathcal{D})  = F^q_{1, d; 1-\alpha_{L}} \mbox{tr}\{(L + \frac{1}{\tau^2}I_q)^{-1}\}$ & \\
\textbf{Robustness} \\
$MSE(D_S)$-optimality\tnote{3} & $\phi_{MSE(D_S)}(\mathcal{D}) = \exp\{E[\log |\mbox{MSE}(\hat{\boldsymbol{\beta}}_1)|]\}$  & \code{criteria.mseD} \\
& & \code{criteria.mseP}, \\
$MSE(L)$-optimality & $\phi_{MSE(L)}(\mathcal{D}) = \sigma^2\mbox{trace}[(X^\top X)^{-1} + \tau^2 AA^\top]$& \code{criteria.mseL} \\ 
\hline \\
\end{tabular}
\begin{tablenotes}
       \item [1] Implemented in functions \code{criteria.GD}, \code{criteria.GDP}, \code{criteria.mseD} and \code{criteria.mseP}. 
       \item [2] Implemented in functions \code{criteria.GL}, \code{criteria.GLP} and \code{criteria.mseL}. 
       \item [3] With the expected log-determinant approximated via Monte Carlo sampling (\code{criteria.mseD}) or evaluated for a point prior (\code{criteria.mseP}).
\end{tablenotes}
\end{center}
\caption{Design selection criteria implemented in \pkg{MOODE}\label{tab:criteria}}
\end{threeparttable}
\end{table}

\subsection{Model estimation and inference}\label{sec:model-inf}

The most common design selection criteria aim at estimation or inference for the primary model~\eqref{eq:primary_model}. For estimation of $\boldsymbol{\beta}_1$ when $\sigma^2$ is known, or for sufficiently large experiments, \pkg{MOODE} includes functions implementing $D_S$- and $L$-optimality that minimise the following functions of design $\mathcal{D}$, treating the intercept $\beta_0$ as a nuisance parameter: 
\begin{align*}
\phi_{D_S}(\mathcal{D}) & =  \left|\left[X_1^\top\left(I_n - \frac{1}{n}J_n\right)X_1\right]^{-1}\right|\,, & \quad \mbox{($D_S$-optimality)} \\
\phi_L(\mathcal{D}) & =  \mbox{tr}\left\{L^\top\left(X_1^\top(I_n - \frac{1}{n}J_n)X_1\right)^{-1}L\right\}\,. & \quad \mbox{($L$-optimality)}
\end{align*}
Here $I_n$ and $J_n$ are the $n\times n$ identity and all-ones matrices, respectively, and $L$ is a $p_1 \times p$ matrix defining $p_1$ linear combinations $L^\top\boldsymbol{\beta}_1$ of interest in the experiment. In \pkg{MOODE}, the default choice is $LL^\top$ being a $p$-dimensional diagonal matrix with entries 1 except if the corresponding $\beta_j$ corresponds to a quadratic predictor, when the value is 0.25. This choice leads to a weighted $A$-optimality criterion with quadratic effects having the same weight in the computation of the average variance as linear and interaction effects; see \citet{achj1993} and \citet{GilmourandTrinca2012}.

Inference for model~\eqref{eq:primary_model} also depends on the availability of an unbiased estimator for $\sigma^2$, such as the pure error estimate obtained from fitting model~\eqref{eq:full_treat_model}. To ensure replication is available in the experiment to provide such an estimate, \citet{GilmourandTrinca2012} suggested a class of criteria that explicitly incorporate the F-distribution quantiles on which parameter confidence regions depend. In particular, they defined $(DP)_S$- and $LP$-optimal designs that minimise
\begin{align*}
\phi_{(DP)_S}(\mathcal{D}) & =  F_{p,d;1-\alpha}^p\phi_{D_s}(\mathcal{D})\,, & \quad \mbox{($(DP)_S$-optimality)} \\
\phi_{LP}(\mathcal{D}) & =  F_{1,d;1-\alpha}\phi_L(\mathcal{D})\,, & \quad \mbox{($LP$-optimality)}
\end{align*}
where $d = n-t$ is the number of replicated treatments in the experiment, $\alpha$ is a pre-chosen significance level and $F_{df1, df2; 1-\alpha}$ is the quantile of an F-distribution with $df1$ and $df2$ degrees of freedom such that the probability of being less than or equal to this quantile is $1-\alpha$. 

\subsection{Model sensitivity}\label{sec:model-sens}

The alternative experimental aim of model sensitivity is concerned with determining lack-of-fit in the direction of the potential terms. It is common to derive criteria for design assessment under model sensitivity from the non-centrality parameter for an F-test for $\boldsymbol{\beta}_2 = \boldsymbol{0}_q$ (see, e.g., \citealp{Atkinson1975Design,Goosetal2005}).

We make use of a related quantity derived from the posterior variance-covariance matrix for $\boldsymbol{\beta}_2$, conditional on the value of $\sigma^2$,
\begin{align*}
\Sigma_2 & = \sigma^2\left\{X_2^\top\left[I_q - X(X^\top X)^{-1}X^\top\right]X_2 + \frac{1}{\tau^2}I_q\right\}^{-1} \\
& = \sigma^2\left(R + \frac{1}{\tau^2}I_q \right)^{-1}\,,
\end{align*}
where we assume normally-distributed errors in model~\eqref{eq:primary_potential_model} and independent priors motivated by \citet{DuMouchel1994} and \citet{Kobilinsky1998}: non-informative priors for $\beta_0$ and $\boldsymbol{\beta}_1$ with variance tending to infinity; and a normal distribution for $\boldsymbol{\beta}_2\sim \mathcal{N}\left(\boldsymbol{0}_q, \sigma^2\tau^2I_q\right)$ for $\tau^2>0$.

The ability of the design to make inference about the potential terms, and hence detect any lack of fit in the direction of model~\eqref{eq:primary_potential_model} can be quantified via functionals of $R + \frac{1}{\tau^2}I_q$. We define Lack-of-fit $DP$- and $LP$-criteria that minimise
\begin{align*}
\phi_{LoF-DP}(\mathcal{D}) & = F^q_{q, d; 1-\alpha_{L}} \left|R + \frac{1}{\tau^2}I_q\right|^{-1}\,, & \quad (\mbox{LoF-$DP$-optimality)} \\ 
\phi_{LoF-LP}(\mathcal{D}) & = F_{1, d; 1-\alpha_{L}} \mbox{tr}\left\{L^\top\left(R + \frac{1}{\tau^2}I_q\right)^{-1}L\right\}\,. & \quad (\mbox{LoF-$LP$-optimality)}
\end{align*}
The inclusion of the $F$-quantiles recognises the need for a pure error estimate for $\sigma^2$ and encourages replication of treatments. Both criteria target designs with matrices $X_1$ and $X_2$ being (near) orthogonal to each other, which will also maximise the power of the lack-of-fit test for the potential terms.

\subsection{Model robustness}\label{sec:model-robust}

Complementary to detecting lack of fit, it is also desirable to be able to estimate $\boldsymbol{\beta}_1$ from model~\eqref{eq:primary_model} protected from contamination from the potential terms. Assuming least squares estimation, or equivalently maximum likelihood estimation with normal errors, it is natural to consider the mean squared error (MSE) of $\hat{\boldsymbol{\beta}}_1$ which can be characterised via the MSE matrix \citep{FedorovMontepiedra1997}:
\begin{align}
\label{eq::MSE}
\mbox{MSE}\left(\hat{\boldsymbol{\beta}}_1\right)& = \mathtt{E}_{\boldsymbol{Y}}[(\hat{\boldsymbol{\beta}}_1 -\boldsymbol{\beta}_1)(\hat{\boldsymbol{\beta}}_1 - \boldsymbol{\beta})_1^\top]\notag\\
& = \sigma^2[X_1^\top (I_n - \frac{1}{n}J_n) X_1]^{-1} + A_1\boldsymbol{\beta}_2\boldsymbol{\beta}_2^\top A_1^\top\,, 
\end{align}
where 
$$
A_1 = \left[X_1^\top \left(I_n - \frac{1}{n}J_n\right) X_1\right]^{-1}X_1^\top \left(I_n - \frac{1}{n}J_n\right)X_2
$$ 
is the $p\times q$ alias matrix between the primary and potential terms (excluding the intercept); see \citet{EgorovaThesis2017} for the derivation of the MSE matrix.

An analogy of variance-based alphabetic criterion is to consider functionals of this matrix. For the determinant, letting $M = X_1^\top (I_n - \frac{1}{n}J_n) X_1$ and $\tilde{\boldsymbol{\beta}}_2 = \boldsymbol{\beta}_2 / \sigma$, we obtain
\begin{align*}
\left|\mbox{MSE}\left(\hat{\boldsymbol{\beta}}_1\right)\right| & = \sigma^{2p}\left| M^{-1} + A_1\tilde{\boldsymbol{\beta}}_2\tilde{\boldsymbol{\beta}}_2^\top A_1^\top\right| \\
& = \sigma^{2p}\left|M^{-1}\right|\left(1 + \tilde{\boldsymbol{\beta}}_2^\top X^\top_2X_1M^{-1}X_1^\top X_2\tilde{\boldsymbol{\beta}}_2\right)\,,
\end{align*}
using the matrix determinant lemma \citep[p.~417]{Harville2006matrix}. Hence, on a log scale,
\begin{equation}\label{eq:log-det-mse}
\log \left|\mbox{MSE}\left(\hat{\boldsymbol{\beta}}_1\right)\right| = p\log \sigma^2 + \log \left| M^{-1} \right| + \log\left(1 + \tilde{\boldsymbol{\beta}}_2^\top X^\top_2X_1M^{-1}X_1^\top X_2\tilde{\boldsymbol{\beta}}_2\right)\,.
\end{equation}
The first summand is constant with respect to the design, and hence can be excluded from any objective function. The second summand is the (log-scale) $D_S$-optimality objective function, focused on precise estimation of the primary terms $\boldsymbol{\beta}_1$. And the third summand quantifies the bias introduced by not including the potential terms in the fitted model. 

The trade-off between variance and bias is clearly controlled by the relative values of $\sigma^2$ and $\boldsymbol{\beta}_2$; in particular, equation~\eqref{eq:log-det-mse} is reduces to a form equivalent to the $D_S$-optimality objective function if $\boldsymbol{\beta}_2 = \boldsymbol{0}_q$. 
In general, the values of $\boldsymbol{\beta}_2$ will not be known. Using the same prior assumed in Section~\ref{sec:model-sens} we obtain $\tilde{\boldsymbol{\beta}}_2\sim \mathcal{N}\left(\boldsymbol{0}_q, \tau^2I_q\right)$, and the expectation of~\eqref{eq:log-det-mse} can be approximated using Monte Carlo simulation as
\begin{align}
E\left[\log \left|\mbox{MSE}\left(\hat{\boldsymbol{\beta}}_1\right)\right|\right] & = 
p\log \sigma^2 + \log \left| M^{-1} \right| + E\left[\log\left(1 + \tilde{\boldsymbol{\beta}}_2^\top X^\top_2X_1M^{-1}X_1^\top X_2\tilde{\boldsymbol{\beta}}_2\right)\right] \nonumber\\
& \approx p\log \sigma^2 + \log \left| M^{-1} \right| + \frac{1}{B}\sum_{i=1}^B\log\left(1 + \tilde{\boldsymbol{\beta}}_{2i}^\top X^\top_2X_1M^{-1}X_1^\top X_2\tilde{\boldsymbol{\beta}}_{2i}\right)\,,\label{eq:mc-mse}
\end{align}
where $\tilde{\boldsymbol{\beta}}_{21}, \ldots,\tilde{\boldsymbol{\beta}}_{2B}$ are a sample from the prior distribution.

Obtaining a precise approximation via~\eqref{eq:mc-mse} may require large values of $B$ and hence be computationally expensive. As an alternative, we can take a ``locally optimal'' approach \citep{chernoff} and choose a point prior for $\tilde{\boldsymbol{\beta}}_2$ at which to evaluate~\eqref{eq:log-det-mse}. One possibility is to set $\boldsymbol{\beta}_2 = \pm \sigma\tau \boldsymbol{1}_q$, and hence $\tilde{\boldsymbol{\beta}}_2 = \pm \tau \boldsymbol{1}_q$. This choice fixes each potential parameter to be one standard deviation from the prior mean. Taking $\tilde{\boldsymbol{\beta}}_2 = \tau \boldsymbol{1}_q$, without loss of generality, we obtain
\begin{align}
E\left[\log \left|\mbox{MSE}\left(\hat{\boldsymbol{\beta}}_1\right)\right|\right] & \approx 
p\log \sigma^2 + \log \left| M^{-1} \right| + \log\left(1 + \tau^2\boldsymbol{1}_q^\top X^\top_2X_1M^{-1}X_1^\top X_2\boldsymbol{1}_q\right)\,.\label{eq:pp-mse} 
\end{align}
We define the $MSE(D_S)$-criterion via minimisation of
$$
\phi_{MSE(D_S)}(\mathcal{D}) = \exp\left\{E\left[\log \left|\mbox{MSE}\left(\hat{\boldsymbol{\beta}}_1\right)\right|\right]\right\}\,,
$$
with the expected log determinant approximated using either~\eqref{eq:mc-mse} or~\eqref{eq:pp-mse}.

We can also consider an $MSE(L)$-criterion formed from the trace of the MSE matrix. As a linear functional, the expectation can be found directly:
\begin{align*}
\phi_{MSE(L)}(\mathcal{D}) = E\left\{\mbox{trace}\left[\mbox{MSE}\left(\hat{\boldsymbol{\beta}}_1\right)\right]\right\} & = \mbox{trace}\left\{E\left[\mbox{MSE}\left(\hat{\boldsymbol{\beta}}_1\right)\right]\right\} \\
& = \mbox{trace}\left[\sigma^2M^{-1} + E\left(A_1\boldsymbol{\beta}_2\boldsymbol{\beta}_2^\top A_1^\top\right)\right] \\
& = \sigma^2\mbox{trace}\left[M^{-1} + \tau^2 A_1^\top A_1\right]\,.
\end{align*}

\subsection{Compound criteria}\label{sec:compound}

The main contribution of the \pkg{MOODE} package is the ability to find optimal designs under a combination of the criteria for estimation and inference, model sensitivity and model robustness described above. Criteria objective functions are combined via a product of individual objective functions, each raised to the power of a non-negative weight (or, equivalently, a product of weighted efficiencies). Table~\ref{tab:comp-crit} summarises the compound criteria that are available in the package. In addition, the \textit{Generalised} $D$- and $A$-criteria from \citet{Goosetal2005} are also available via specification of \code{criterion.choice = "GD"} or \code{criterion.choice = "GL"} when creating a \code{mood} object. 

\begin{table}
\centering
\begin{threeparttable}
\begin{tabular}{lcccccccccc}
 & \multicolumn{8}{c}{Individual criteria} \\
 \hline
 \multicolumn{9}{l}{\textbf{Generalized criteria}: not accounting for pure error estimation} \Bstrut\Tstrut \\
 & $D_S$ & $L$ & LoF-$D$ & LoF-$L$ & Bias-$D$ & Bias-$L$ \\
\hline
\texttt{GD} & \checkmark & & \checkmark & & \checkmark &  \\
\texttt{GL} & & \checkmark & & \checkmark & & \checkmark \\
\hline
\multicolumn{9}{l}{\textbf{Amended Generalized criteria}: accounting for pure error estimation } \Bstrut\Tstrut \\

 & $D_S$ & $L$ & $(DP)_S$ & $LP$ & LoF-$DP$ & LoF-$LP$ & Bias-$D$ & Bias-$L$  \\
 \hline
\texttt{GDP} & \checkmark & & \checkmark & & \checkmark &  
 & \checkmark &\\
\texttt{GLP} & & \checkmark & & \checkmark & & \checkmark  & & \checkmark\\ \hline
\multicolumn{9}{l}{\textbf{Compound MSE criteria}} \Bstrut\Tstrut \\
 & $(DP)_S$ & $LP$ & LoF-$DP$ & LoF-$LP$  & $MSE(D_S)$ & $MSE(L)$ \\ \hline
\texttt{MSE.D}\tnote{1}  & \checkmark & & \checkmark  &   & \checkmark & \\
\texttt{MSE.P}\tnote{2} & \checkmark & & \checkmark  &   & \checkmark & \\
\texttt{MSE.L} & & \checkmark & & \checkmark &  & \checkmark \\
\hline
\end{tabular}
\begin{tablenotes}
       \item [1] With the expected log-determinant approximated via Monte Carlo sampling 
       \item [2] With the expected log-determinant evaluated for a point prior.
\end{tablenotes}
\caption{Component criteria (columns) that can be combined in the \pkg{MOODE} package via argument \code{criterion.choice} to function \pkg{mood} (rows).}
\label{tab:comp-crit}
\end{threeparttable}
\end{table}

Here we focus on determinant- and trace-based compound criteria, defined as
\begin{equation}\label{eq:MSE-compound-D}
\phi_{det}(\mathcal{D}) = \phi_{(DP)_S}(\mathcal{D})^{\kappa_{DP}}\times \phi_{LoF-DP}(\mathcal{D})^{\kappa_{LoF-DP}} \times \phi_{MSE(D_S)}(\mathcal{D})^{\kappa_{MSE(D)}}
\end{equation}
and
\begin{equation}\label{eq:MSE-compound-L}
\phi_{trace}(\mathcal{D}) = \phi_{LP}(\mathcal{D})^{\kappa_{LP}}\times \phi_{LoF-LP}(\mathcal{D})^{\kappa_{LoF-LP}} \times \phi_{MSE(L)}(\mathcal{D})^{\kappa_{MSE(L)}}\,,
\end{equation}
respectively, with all weights $\kappa_\star\ge 0$, $\kappa_{DP} + \kappa_{LoF-DP} + \kappa_{MSE(D)} = 1$ and $\kappa_{LP} + \kappa_{LoF-LP} + \kappa_{MSE(L)} = 1$. 

These criteria are available via specification of the argument \code{criterion.choice = "MSE.D"} or \code{criterion.choice = "MSE.P"} (for the determinant-based criterion with, respectively, the expected log-determinant approximated via Monte Carlo sampling or evaluated for a point prior), or \code{criterion.choice = "MSE.L"} (for the trace-based criterion). The following code specifies the determinant-based criterion for our earlier example.

\begin{CodeInput}
ex.mood <- mood(K = 2, Levels = 3, Nruns = 24, 
                model_terms = list(primary.terms = c("x1", "x2"), 
                                   potential.terms = c("x12", "x22")), 
                criterion.choice = "MSE.D", 
                control = list(Biter = 1000), 
                kappa = list(kappa.DP = 1 / 3, kappa.LoF = 1 / 3, 
                kappa.mse = 1 / 3))
\end{CodeInput}

The default number of Monte Carlo samples is $B = 50$ in~\eqref{eq:mc-mse}, which we can change via the \code{control} argument. We specify criteria weights via the \code{kappa} argument.

\section{Design search algorithms}\label{sec:algorithms}

Within the \pkg{MOODE} package, the function \code{Search} finds efficient compound designs using exchange algorithms. Two algorithms are implemented, a modified-Fedorov point exchange algorithm (\citealp{CookandNachtsheim1980}; the default for $k<=4$ factors) and a coordinate exchange algorithm (\citealp{MeyerNachtsheim1995}; the default for $k>=5$ factors). The algorithm can also be specified via the argument \code{algorithm = "ptex"} (point exchange) or \code{algorithm = "coordex"} (coordinate exchange). Details of the of the design problem such as number of factors and units, models and criterion are passed via an object generated by the \code{mood} function, as exemplified in Section~\ref{sec:compound}. Below, if the argument \code{algorithm} is not specified, it defaults to \code{algorithm = "ptex"}, as $k=2$ in \code{ex.mood}. To use the coordinate exchange algorithm in this example, we must specify \code{algorithm = "coordex"}.

\begin{CodeInput}
search.ex <- Search(ex.mood)
search.ex <- Search(ex.mood, algorithm = "coordex")
\end{CodeInput}

These exchange algorithms are heuristics, and convergence to the optimal solution is not guaranteed. Hence, multiple instances of each algorithm should be performed, starting at different randomly chosen starting designs, to help avoid local optima. In \pkg{MOODE}, the number of starts is specified as a \code{control} argument, \code{Nstarts}, to the \code{mood} function, with default 10. Depending on the consistency of the results and the complexity of the assumed models and multi-objective criteria, we generally recommend increasing this number substantially.

Running these separate instances in parallel, using different processors on a single computer, is implemented using the \pkg{doFuture} package \citep{RJ-2021-048} and by setting the argument \code{parallel = T}. 

\begin{CodeInput}
library(doFuture)
plan(multisession)
search.ex <- Search(ex.mood, parallel = T)
plan(sequential)    
\end{CodeInput}

By default, a \pkg{doFuture} multisession plan uses all available processors. The number to use can instead be specified by the \code{workers} argument to \code{plan}.

A \code{Search} object contains a variety of output, including the final design (\code{X.design}), the model matrices for the primary (\code{X1}) and potential (\code{X2}) terms, the final value of the compound objective function (\code{compound.value}) and the objective function values for each of the \code{Nstarts} tries of the algorithm (\code{path}).

\begin{CodeInput}
fd <- search.ex$X.design[order(search.ex$X1[, 1]),]
cbind(fd[1:12, ], fd[13:24, ])
\end{CodeInput}   

\begin{CodeOutput}
      x1 x2 x1 x2
 [1,] -1 -1  0  0
 [2,] -1 -1  0  1
 [3,] -1 -1  0  1
 [4,] -1 -1  1 -1
 [5,] -1  0  1 -1
 [6,] -1  0  1 -1
 [7,] -1  1  1 -1
 [8,] -1  1  1  0
 [9,] -1  1  1  0
[10,] -1  1  1  1
[11,]  0 -1  1  1
[12,]  0 -1  1  1
\end{CodeOutput}

\begin{CodeInput}
search.ex$path
\end{CodeInput}   

\begin{CodeOutput}
[1] 0.1889336 0.1891478 0.1875964 0.1891478 0.1891478 0.1875964 0.1875964 
0.1889324 0.1891478 0.1875964
\end{CodeOutput}

\section[sec4]{Case studies}
\label{sec:examples}

In this section, we use two case studies to demonstrate finding multi-objective optimal designs under model contamination using the \pkg{MOODE} package: (i) an experiment based on the case study from \citet{EgorovaGilmour2022} with three factors with $n = 36$ runs; and (ii) designs with $n=12$ runs to investigate between three and 9 factors, comparing to the classic Plackett-Burman design \citep{pb}.

\subsection{Egorova and Gilmour case study}
\label{sec:eg}

The case study in \citet{EgorovaGilmour2022} considered a block design for a animal food supplement study; here, we find a completely randomised design in the same number of runs. The determinant-based compound criterion~\eqref{eq:MSE-compound-D} is applied, combining $(DP)_S$-optimality (for inference using a primary model), lack-of-fit (in the direction of a potential model) and the $MSE(D_S)$ criterion (for estimation of a primary model robust to additional terms from a potential model). We investigate the impact of the choice of different weights for these three criteria as defined in the following code chunk, where columns 1, 2 and 3 of the \code{kappa} matrix specify $\kappa_{DP}$, $\kappa_{LoF-DP}$ and $\kappa_{MSE(D)}$ respectively.

\begin{CodeInput}
set.seed(16092024)
library(MOODE)
kappa <- matrix(
  c(1/3, 1/3, 1/3,
    0.4, 0.2, 0.4,
    0.25, 0.25, 0.5,
    1, 0, 0,
    0, 1, 0,
    0, 0, 1),
  ncol = 3, byrow = T)
\end{CodeInput}

Following \citet{EgorovaGilmour2022}, the primary and potential models are, respectively, a full second-order model including linear, quadratic and two-way interaction terms ($p=9$), and an encompassing model that also includes cubic terms, linear-by-linear-by-linear and quadratic-by-linear interactions ($q=10$). These models can be straightforwardly specified using the \code{model.terms} argument of the \code{mood} function.

To efficiently find the six designs defined by the rows of \code{kappa} we employ the parallel functionality of the \code{Search} function. 

\begin{CodeInput}
designs_eg <- list()
eg_cs <- list()
plan(multisession)
for(i in 1:nrow(kappa1)) {
  eg_cs[[i]] <- mood(K = 3, Levels = 5, Nruns = 36, 
                     criterion.choice = "MSE.P", 
                     kappa = list(kappa.DP = kappa1[i, 1], 
                                  kappa.LoF = kappa1[i, 2], 
                                  kappa.mse = kappa1[i, 3]), 
                     model_terms = list(primary.model = "second_order", 
                                        potential.model = 
                                          c("cubic_terms", 
                                            "third_order_terms")),
                     control = list(Biter = 50))
  Search_eg_cs <- Search(eg_cs[[i]], parallel = TRUE)
  designs_eg[[i]] <- Search_eg_cs
}
plan(sequential)
\end{CodeInput}

We loop through the \code{kappa} matrix and use the rows to specify the weights (\code{kappa.DP}, \code{kappa.LoF} and \code{kappa.mse}) for six different compound criteria. For $k=3$ factors, a point exchange algorithm is used with a candidate list constructed using \code{Levels = 5} values for each factor. The resulting designs are saved in a list. To evaluate objective function~\eqref{eq:mc-mse} for the $MSE(D_S)$ component, we use $B = 1000$ Monte Carlo samples. Efficiencies of the resulting designs are given in Table~\ref{tab:egex}.

\begin{table}
\centering
\begin{tabular}{rrrrrrrr}
\toprule
\multicolumn{3}{c}{ } & \multicolumn{3}{c}{Efficiency (\%)} & \multicolumn{2}{c}{DF} \\
\cmidrule(l{3pt}r{3pt}){4-6} \cmidrule(l{3pt}r{3pt}){7-8}
$\kappa_{DP}$ & $\kappa_{LoF-DP}$ & $\kappa_{MSE(D)}$ & $(DP)_S$ & LoF-$DP$ & $MSE(D_S)$ & PE & LoF\\
\midrule
0.33 & 0.33 & 0.33 & 89.11 & 100.54 & 95.84 & 16 & 10\\
0.40 & 0.20 & 0.40 & 94.20 & 89.30 & 99.50 & 17 & 9\\
0.25 & 0.25 & 0.50 & 90.71 & 94.26 & 99.84 & 15 & 11\\
1.00 & 0.00 & 0.00 & 100.00 & 67.05 & 98.73 & 22 & 4\\
0.00 & 1.00 & 0.00 & 77.47 & 100.00 & 90.35 & 14 & 12\\
0.00 & 0.00 & 1.00 & 73.08 & 80.31 & 100.00 & 9 & 17\\
\bottomrule
\end{tabular}
\caption{\label{tab:egex}Efficiencies under individual criteria and degrees of freedom for the Egorova and Gilmour case study designs defined by six different sets of criteria weights.}
\end{table}

The $(DP)_S$-, LoF-$DP$- and $MSE(D_S)$-optimal designs (last three rows of Table~\ref{tab:egex}) provide benchmarks for the three different compound criteria. Note that the empirical nature of the design optimisation is reflected in an efficiency value slightly greater than 100\% (100.54) under the LoF criterion for the first compound design. All three designs obtained from proper compound criteria provide good compromises across the three individual criteria, having high efficiencies for inference under an assumed model, detection of lack-of-fit and estimation under a misspecified model. These trade offs are also reflected in the balances achieved between pure error and lack-of-fit degrees of freedom. Designs from the individual criteria each have poor efficiency under at least one of the other two criteria. The design obtained using $\kappa_{DP}=0.4$, $\kappa_{LoF-DP} = 0.2$ and $\kappa_{MSE(D)}=0.4$ achieves a good trade-off between objectives and is given in Table~\ref{tab:egexdes}, alongside the $(DP)_S$- and $MSE(D_S)$-optimal designs. The compound design (with 19 treatments) compromises between the replication found in the $(DP)_S$ (14 treatments) and $MSE(D_S)$ designs (27 treatments). Note that treatments 114 (with $x_3$ set to 0.5) and 56 (with $x_2$ set to 0.5) are included in the compound and $MSE(D)$ designs, which are unusual choices given the primary model. It is possible these design could be improved further do using more random starts of the algorithm.

\begin{table}
\centering
\begin{tabular}{rrrrrrrrrrrr}
\toprule
\multicolumn{4}{c}{Compound} & \multicolumn{4}{c}{$(DP)_S$-optimal} & \multicolumn{4}{c}{$MSE(D_S)$-optimal} \\
\cmidrule(l{3pt}r{3pt}){1-4} \cmidrule(l{3pt}r{3pt}){5-8} \cmidrule(l{3pt}r{3pt}){9-12}
Trt label & $x_{1}$ & $x_{2}$ & $x_{3}$ & Trt label & $x_{1}$ & $x_{2}$ & $x_{3}$ & Trt label & $x_{1}$ & $x_{2}$ & $x_{3}$\\
\midrule
1 & -1 & -1 & -1.0 & 1 & -1 & -1 & -1 & 1 & -1 & -1.0 & -1\\
1 & -1 & -1 & -1.0 & 1 & -1 & -1 & -1 & 1 & -1 & -1.0 & -1\\
1 & -1 & -1 & -1.0 & 1 & -1 & -1 & -1 & 3 & -1 & -1.0 & 0\\
5 & -1 & -1 & 1.0 & 5 & -1 & -1 & 1 & 5 & -1 & -1.0 & 1\\
5 & -1 & -1 & 1.0 & 5 & -1 & -1 & 1 & 5 & -1 & -1.0 & 1\\
5 & -1 & -1 & 1.0 & 5 & -1 & -1 & 1 & 11 & -1 & 0.0 & -1\\
11 & -1 & 0 & -1.0 & 13 & -1 & 0 & 0 & 13 & -1 & 0.0 & 0\\
13 & -1 & 0 & 0.0 & 13 & -1 & 0 & 0 & 15 & -1 & 0.0 & 1\\
13 & -1 & 0 & 0.0 & 13 & -1 & 0 & 0 & 21 & -1 & 1.0 & -1\\
21 & -1 & 1 & -1.0 & 21 & -1 & 1 & -1 & 21 & -1 & 1.0 & -1\\
21 & -1 & 1 & -1.0 & 21 & -1 & 1 & -1 & 23 & -1 & 1.0 & 0\\
23 & -1 & 1 & 0.0 & 21 & -1 & 1 & -1 & 25 & -1 & 1.0 & 1\\
25 & -1 & 1 & 1.0 & 25 & -1 & 1 & 1 & 25 & -1 & 1.0 & 1\\
25 & -1 & 1 & 1.0 & 25 & -1 & 1 & 1 & 51 & 0 & -1.0 & -1\\
51 & 0 & -1 & -1.0 & 25 & -1 & 1 & 1 & 53 & 0 & -1.0 & 0\\
53 & 0 & -1 & 0.0 & 53 & 0 & -1 & 0 & 55 & 0 & -1.0 & 1\\
53 & 0 & -1 & 0.0 & 53 & 0 & -1 & 0 & 56 & 0 & -0.5 & -1\\
65 & 0 & 0 & 1.0 & 53 & 0 & -1 & 0 & 63 & 0 & 0.0 & 0\\
65 & 0 & 0 & 1.0 & 61 & 0 & 0 & -1 & 65 & 0 & 0.0 & 1\\
71 & 0 & 1 & -1.0 & 61 & 0 & 0 & -1 & 71 & 0 & 1.0 & -1\\
71 & 0 & 1 & -1.0 & 61 & 0 & 0 & -1 & 73 & 0 & 1.0 & 0\\
75 & 0 & 1 & 1.0 & 75 & 0 & 1 & 1 & 75 & 0 & 1.0 & 1\\
101 & 1 & -1 & -1.0 & 75 & 0 & 1 & 1 & 75 & 0 & 1.0 & 1\\
101 & 1 & -1 & -1.0 & 101 & 1 & -1 & -1 & 101 & 1 & -1.0 & -1\\
105 & 1 & -1 & 1.0 & 101 & 1 & -1 & -1 & 101 & 1 & -1.0 & -1\\
105 & 1 & -1 & 1.0 & 101 & 1 & -1 & -1 & 103 & 1 & -1.0 & 0\\
105 & 1 & -1 & 1.0 & 105 & 1 & -1 & 1 & 105 & 1 & -1.0 & 1\\
111 & 1 & 0 & -1.0 & 105 & 1 & -1 & 1 & 105 & 1 & -1.0 & 1\\
111 & 1 & 0 & -1.0 & 115 & 1 & 0 & 1 & 111 & 1 & 0.0 & -1\\
114 & 1 & 0 & 0.5 & 115 & 1 & 0 & 1 & 113 & 1 & 0.0 & 0\\
121 & 1 & 1 & -1.0 & 121 & 1 & 1 & -1 & 115 & 1 & 0.0 & 1\\
121 & 1 & 1 & -1.0 & 121 & 1 & 1 & -1 & 121 & 1 & 1.0 & -1\\
123 & 1 & 1 & 0.0 & 123 & 1 & 1 & 0 & 121 & 1 & 1.0 & -1\\
123 & 1 & 1 & 0.0 & 123 & 1 & 1 & 0 & 123 & 1 & 1.0 & 0\\
125 & 1 & 1 & 1.0 & 125 & 1 & 1 & 1 & 125 & 1 & 1.0 & 1\\
125 & 1 & 1 & 1.0 & 125 & 1 & 1 & 1 & 125 & 1 & 1.0 & 1\\
\bottomrule
\end{tabular}
\caption{\label{tab:egexdes} Compound optimal design for the Egorova-Gilmour case study found with $\kappa_{DP} = 0.4$, $\kappa_{LoF-DP} = 0.2$ and $\kappa_{MSE(D)} = 0.4$.}
\end{table}

\subsection{A comparison with Plackett-Burman designs}
\label{sec:pb}

The 12-run Plackett-Burman design is perhaps the most widely used, and studied, non-regular fractional factorial design. It can accommodate up to 11 two-level factors for the orthogonal estimation of main effects. The main effect estimator for each factor is biased by all two-factor interactions not involving that factor, with aliasing coefficients (entries of $A_1$) given by $\pm 1/3$.

Here, we use the \pkg{MOODE} package to find alternative two-level designs for $k=3,\ldots,9$ factors, using the trace-based compound criterion that minimizes~\eqref{eq:MSE-compound-L}. The below code chunk defines the values of $\kappa_{LP}$ (\code{kappa} column 1), $\kappa_{LoF-LP}$ (column 2) and $\kappa_{MSE(L)}$ (column 3) for which designs are sought.

\newpage

\begin{CodeInput}
set.seed(10122024)
kappa2 <- matrix(
  c(1/3, 1/3, 1/3,
    0.25, 0.25, 0.5,
    1, 0, 0,
    0, 1, 0,
    0, 0, 1,
    0, 0, 0),
  ncol = 3, byrow = T)    
\end{CodeInput}

The last four rows of \code{kappa} define the individual $LP$-, LoF-$LP$-, $MSE(L)$ and $L$-criteria. The primary model considered consists of all $k$ main effects, corresponding to typical use of a Plackett-Burman design, with the potential model also including all two-factor interactions. The following code uses a double loop to employ the \code{Search} function to find designs for all these choices of $k$ and the $\kappa$ weights. When the row sum of \code{kappa} is equal to zero, the \code{pb} function from the \pkg{FrF2} package is used to obtain a Plackett-Burman design, which will be $L$-optimal for a main effects model.

\begin{CodeInput}
designs_pb <- list()
mood_pb <- list()
library(FrF2)
plan(multisession)
for(j in 3:9) {
  temp <- list()
  pb <- list()
  for(i in 1:nrow(kappa2)) {
    if(sum(kappa2[i, ]) == 0) { 
      X1 <- undesign(pb(nruns = 12, nfactors = j)) |>
        mutate(across(all_of(1:j), ~ 2 * as.numeric(.x) - 3)) |>
        mutate(trt = 1:12, intercept = rep(1, 12), .before = A)
      temp[[i]] <- list(X1 = as.matrix(X1))
      temp[[i]]$X2 <- model.matrix(~ (.)^2, X1)[, -(1:(j + 1))]
    } else { 
      pb[[i]] <- mood(K = j, Levels = 2, Nruns = 12, 
                      criterion.choice = "MSE.L",
                      kappa = list(kappa.LP = kappa2[i, 1], 
                                   kappa.LoF = kappa2[i, 2], 
                                   kappa.mse = kappa2[i, 3]), 
                      model_terms = list(primary.model = "main_effects", 
                                         potential.model = 
                                                  "linear_interactions"),                       control = list(Nstarts = 200))
      Search_pb <- Search(pb[[i]], parallel = TRUE, update.info = TRUE, 
                          algorithm = "ptex")
      temp[[i]] <- Search_pb
    }
  }
  designs_pb[[j]] <- temp
  mood_pb[[j]] <- pb
}
plan(sequential)  
\end{CodeInput}

\begin{figure}
    \centering
    \includegraphics[width=0.99\linewidth]{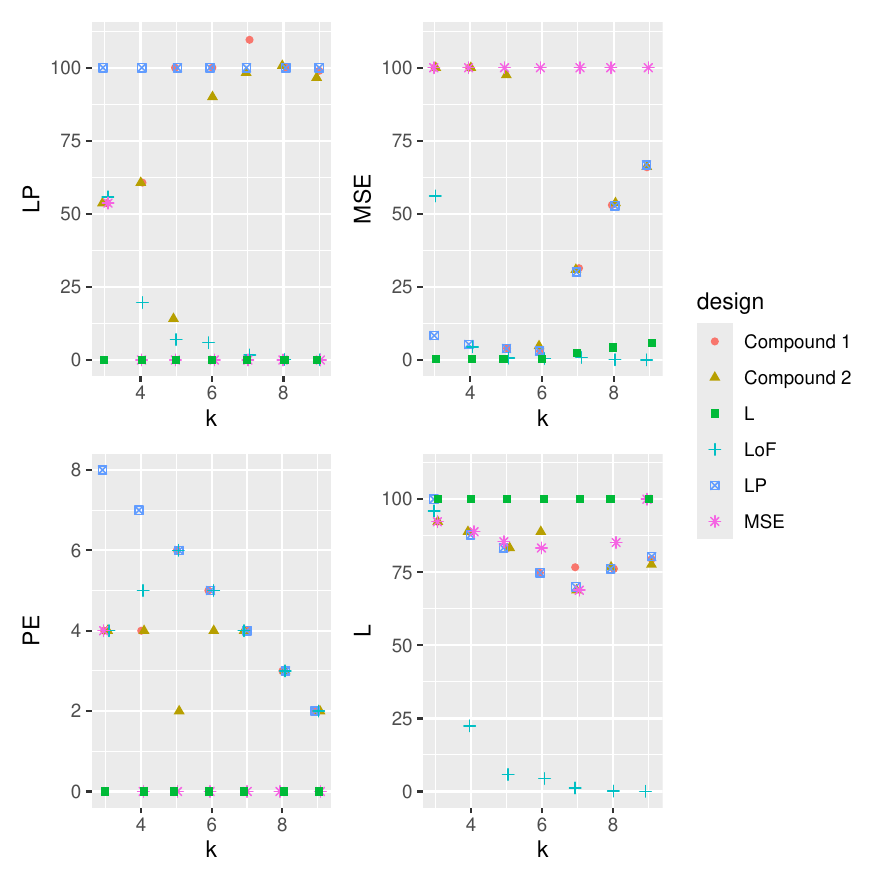}
    \caption{Design efficiencies under the $LP$, $MSE(L)$ and $L$ criteria, along with the pure error degrees of freedom (PE) for two-level designs with $k=3, \ldots,9$ factors and $n = 12$ runs.}
    \label{fig:pb}
\end{figure}

The efficiencies of the resulting designs are displayed in Figure~\ref{fig:pb} under the $LP$-, $MSE(L)$- and $L$-criteria, along with the pure error (PE) degrees of freedom. As noted above, the $L$-optimal designs are simply 12-run Plackett-Burman designs (subsets of columns of the full designs). These designs tend to lack replication, and hence have few or no PE degrees of freedom. This lack of PE degrees of freedom leads to zero efficiency under the $LP$-criterion for larger $k$. For $k>4$, the $MSE(L)$-optimal designs also lack pure error degrees of freedom, leading to zero $LP$-efficiency. The compound designs achieve a minimum of 50\% $LP$-efficiency, and typically achieve somewhat higher, especially for larger numbers of factors. They tend to have four PE degrees of freedom when possible, dropping as the number of factors increases (when the fixed experiment size limits the number of degrees of freedom possible). Also note that for $k=7$, the algorithm found a better design under the $LP$-criterion when searching for the compound design under the first set of weights; this indicates the difficulty sometimes found in finding ``pure error" optimal designs, and could probably be mitigated by more random starts or a more sophisticated choice of starting designs. 

We see more complex patterns under the $MSE(L)$-criterion. When $k$ is smaller, there is very much less replication in the $MSE(L)$-optimal designs than for the $LP$-optimal designs, leading to low $MSE(L)$-efficiency for these latter designs. The compound designs start with very high efficiency for $k=3,4$ but they both maintain replication and PE degrees of freedom for $k=5,6$ leading to much lower efficiency. As $k$ increases, the scope for replication decreases due to the fixed experiment size, and the efficiency of the $LP$-optimal and compound design improves. The $L$-optimal designs perform poorly under the $MSE(L)$-criterion. However, it should be noted that there are multiple choices of the subsets of columns of the Plackett-Burman design, each of which is $L$-optimal but may have quite different performance under the other criteria.

Under $L$-optimality, efficiencies for all designs, excluding the LoF design, lie above 50\%, with highest efficiencies being for $k=3$. The LoF designs tend to perform poorly under all other criteria for $k>3$, and in fact the designs found under $LP$-optimality typically had LoF efficiency near or at 100\%. Conversely, for these limited run sizes and low run-size to number-of-factors ratios, the designs which were optimal under both the $LP$- and LoF-criteria were never found from directly optimizing the LoF criteria.

In general, the F-quantiles in the $LP$-optimality are large for the small PE degrees of freedom possible here. Hence for small values of $k$, where replication is possible, the $LP$-designs are dominated by this quantile, and we see poorer performance of these designs under the $MSE(L)$-criterion. For larger values of $k$, only very limited replication is possible and so the designs, and their performances under different criteria, tend to converge.

The compound design for $\kappa_{LP} = \kappa_{LoF-LP} = \kappa_{MSE(L)} = 1/3$ and $k=4$ is given in Table~\ref{tab:pbdes}, together with the corresponding $LP$- and $MSE(L)$-optimal designs. The relative importance of replication to these three criteria is clear, with the $LP$-optimal design having only 5 distinct points (the minimum number possible), compared to 8 distinct points for the compound design and 12 (obviously the maximum) for the $MSE(L)$-optimal design. The compound design only includes treatments also in the other two designs. None of these designs are orthogonal in the main effects, a property of the Plackett-Burman design that is compromised to obtain either PE degrees of freedom or, in the case of the compound and $MSE(L)$-design, better robustness from the potential terms. These latter two designs achieve orthogonality between the main effects and two-factor interactions, i.e. $A_1$ is a zero matrix. Of course, the relative importance of the aliasing is controlled by the choice of $\tau^2$, the prior variance for the potential terms, which is here set to the default of $\tau^2=1$. 

\begin{table}
\centering
\begin{tabular}{rrrrrrrrrrrrrrr}
\toprule
\multicolumn{5}{c}{Compound 1} & \multicolumn{5}{c}{$LP$-optimal} & \multicolumn{5}{c}{$MSE(L)$-optimal} \\
\cmidrule(l{3pt}r{3pt}){1-5} \cmidrule(l{3pt}r{3pt}){6-10} \cmidrule(l{3pt}r{3pt}){11-15}
Trt label & $x_{1}$ & $x_{2}$ & $x_{3}$ & $x_{4}$ & Trt label & $x_{1}$ & $x_{2}$ & $x_{3}$ & $x_{4}$ & Trt label & $x_{1}$ & $x_{2}$ & $x_{3}$ & $x_{4}$\\
\midrule
2 & -1 & -1 & -1 & 1 & 2 & -1 & -1 & -1 & 1 & 1 & -1 & -1 & -1 & -1\\
2 & -1 & -1 & -1 & 1 & 2 & -1 & -1 & -1 & 1 & 2 & -1 & -1 & -1 & 1\\
3 & -1 & -1 & 1 & -1 & 2 & -1 & -1 & -1 & 1 & 3 & -1 & -1 & 1 & -1\\
5 & -1 & 1 & -1 & -1 & 7 & -1 & 1 & 1 & -1 & 5 & -1 & 1 & -1 & -1\\
5 & -1 & 1 & -1 & -1 & 7 & -1 & 1 & 1 & -1 & 6 & -1 & 1 & -1 & 1\\
8 & -1 & 1 & 1 & 1 & 9 & 1 & -1 & -1 & -1 & 8 & -1 & 1 & 1 & 1\\
9 & 1 & -1 & -1 & -1 & 9 & 1 & -1 & -1 & -1 & 9 & 1 & -1 & -1 & -1\\
12 & 1 & -1 & 1 & 1 & 9 & 1 & -1 & -1 & -1 & 11 & 1 & -1 & 1 & -1\\
12 & 1 & -1 & 1 & 1 & 12 & 1 & -1 & 1 & 1 & 12 & 1 & -1 & 1 & 1\\
14 & 1 & 1 & -1 & 1 & 12 & 1 & -1 & 1 & 1 & 14 & 1 & 1 & -1 & 1\\
15 & 1 & 1 & 1 & -1 & 14 & 1 & 1 & -1 & 1 & 15 & 1 & 1 & 1 & -1\\
15 & 1 & 1 & 1 & -1 & 14 & 1 & 1 & -1 & 1 & 16 & 1 & 1 & 1 & 1\\
\bottomrule
\end{tabular}
\caption{Compound optimal design for $k=4$ two-level factors with $n=12$ runs and $\kappa_{LP} = \kappa_{LoF-LP} = \kappa_{MSE(L)} = 1/3$, along with the corresponding $LP$- and $MSE(L)$-optimal designs.\label{tab:pbdes}}
\end{table}

\section[sec5]{Discussion}
\label{sec:disc}

The perils of assuming one model form for designing experiments have been recognised for at least 65 years, when \citet{BoxandDraper1959} discussed the relative importance of variance and bias in constructing response surface designs. Whilst numerous authors have contributed methodology over the intervening years, \pkg{MOODE} provides the first accessible, open source, implementation of a multi-objective approach to the problem.

Our two examples demonstrate much of the functionality of the package, and how it can be employed. There are further aspects, around candidate list construction, choice of design region etc. that we have not discussed. These can be explored via the package functions and their help files.

The code base for \pkg{MOODE} is reasonably modular and hence is straightforward to expand. Future work will add blocking factors to the optimal designs. Other potential extensions include provision of different optimization algorithms (e.g., nature inspired heuristics; \citealp{Chenetal2014}) or incorporation of different model classes (e.g., so-called hybrid nonlinear models; \citealp{hgmg2020}). 

\bibliography{MOODE}

\appendix

\section{Additional code}

The code to produce the tables and figures in Section~\ref{sec:examples} is reproduced in this appendix. The following additional \proglang{R} packages are used.

\begin{CodeInput}
library("patchwork")
library("ggplot2")
library("dplyr")
library("knitr")
\end{CodeInput}

\subsection{Code for Section 4.1}

\begin{CodeInput}
critvals <- NULL
for(i in 1:nrow(kappa1)) {
  X1 <- designs_eg[[i]]$X1
  X2 <- designs_eg[[i]]$X2
  critvals[[i]] <- criteria.mseD(X1, X2, eg_cs[[1]])  
}

# Table 3 of efficiencies.
vals <- sapply(critvals, \(x) c(x$DP, x$LoF, x$mse, x$df))
vals <- cbind(kappa1, t(vals))
vals <- cbind(vals, 36 - vals[, 7] - 10)
vals[, 4] <- 100 * vals[4, 4] / vals[, 4]
vals[, 5] <- 100 * vals[5, 5] / vals[, 5]
vals[, 6] <- 100 * vals[6, 6] / vals[, 6]
knitr::kable(round(vals, digits = 2), 
             col.names = 
               c("$\\kappa_{DP}$", "$\\kappa_{LoF-DP}$", 
                 "$\\kappa_{MSE-D}$",
                 "DP", "LoF", "MSE", "PE", "LoF"),
             escape = F, booktabs = T,
             linesep = "",
             caption = "Efficiencies under individual criteria and 
             degrees of freedom for the Egorova and Gilmour 
             case study designs defined by six different sets of 
             criteria weights.") |>
  kableExtra::add_header_above(header = c(" " = 3, 
                                          "Efficiency (\\\\%)" = 3, 
                                          "DF" = 2), escape = F)

# Table 4 of example designs, ordered by treatment label
X1 <- designs_eg[[2]]$X1
des1 <- X1[order(X1[, 1]), c(1, 3:5)]
X1 <- designs_eg[[4]]$X1
des2 <- X1[order(X1[, 1]), c(1, 3:5)]
X1 <- designs_eg[[6]]$X1
des3 <- X1[order(X1[, 1]), c(1, 3:5)]
knitr::kable(round(cbind(des1[, 1:4], des2[, 1:4], des3[, 1:4]), 
                   digits = 2), 
             col.names = 
               rep(c("Trt label","$x_{1}$", "$x_{2}$", "$x_{3}$"), 3),
             escape = F, booktabs = T,
             linesep = "",
             caption = "Compound optimal design for the Egorova-Gilmour 
             case study found with $\\kappa_{DP} = 0.4$, 
             $\\kappa_{LoF-DP} = 0.2$ and $\\kappa_{MSE(D)} = 0.4$.") |>
  kableExtra::add_header_above(header = c("Compound" = 4, "$DP$-optimal" = 4, 
                                          "$MSE(D)$-optimal" = 4), escape = F)
\end{CodeInput}

\subsection{Code for Section 4.2}

\begin{CodeInput}
pb_results <- matrix(NA, nrow = nrow(kappa2) * 7, ncol = 9)
count <- 0
for(i in 3:9) {
  for(j in 1:nrow(kappa2)) {
    kappa.vec <- c(1/3, 1/3, 1/3) 
    pb_temp <- mood(K = i, Levels = 2, Nruns = 12, criterion.choice = "MSE.L",
                    kappa = list(kappa.LP = kappa.vec[1], 
                                 kappa.LoF = kappa.vec[2], 
                                 kappa.mse = kappa.vec[3]), 
                    model_terms = list(primary.model = "main_effects", 
                                       potential.model = "linear_interactions"))
    pb_results[count + j, 1] <- i
    pb_results[count + j, 2:4] <- kappa2[j, ]
    X1 <- designs_pb[[i]][[j]]$X1
    X2 <- designs_pb[[i]][[j]]$X2
    critvals <- MOODE:::icriteria.mseL(X1, X2, pb_temp) 
    pb_results[count + j, 5] <- critvals$LP
    pb_results[count + j, 6] <- critvals$LoF
    pb_results[count + j, 7] <- critvals$mse
    pb_results[count + j, 8] <- critvals$df
    pb_results[count + j, 9] <- critvals$L
  }
  count <- count + 6
}
pb_results[, 5] <- 100 * ifelse(pb_results[, 5] == 0, 0, 
rep(pb_results[seq(3, 42, by = 6), 5], rep(6, 7)) / pb_results[, 5]) 
pb_results[, 6] <- 100 * ifelse(pb_results[, 6] == 0, 0, 
rep(pb_results[seq(4, 43, by = 6), 6], rep(6, 7)) / pb_results[, 6]) 
pb_results[, 7] <- 100 * ifelse(pb_results[, 7] == 0, 0, 
rep(pb_results[seq(5, 44, by = 6), 7], rep(6, 7)) / pb_results[, 7]) 
pb_results[, 9] <- 100 * ifelse(pb_results[, 9] == 0, 0,  
((1 / 12) * (rep(3:9, rep(6, 7))) / (rep(4:10, rep(6, 7))))  / pb_results[, 9])
colnames(pb_results) <- c("K", "kappa1", "kappa2", "kappa3", "LP", "LoF", 
                          "MSE", "DF", "L")

pb_results <- data.frame(pb_results)

pb_results <- pb_results |>
  mutate(design = case_when(
    kappa1 == 1/3 ~ "Compound 1",
    kappa1 == 0.25 ~ "Compound 2",
    kappa1 == 1 ~ "LP",
    kappa2 == 1 ~ "LoF",
    kappa3 == 1 ~ "MSE",
    kappa1 + kappa2 + kappa3 == 0 ~ "L"
  ))

# Table 5 of example designs, ordered by treatment number
X11 <- designs_pb[[4]][[1]]$X1
X12 <- designs_pb[[4]][[2]]$X1
X13 <- designs_pb[[4]][[3]]$X1
X15 <- designs_pb[[4]][[5]]$X1
X16 <- designs_pb[[4]][[6]]$X1

X11[order(X11[,1]),]
X12[order(X12[,1]),]
X13[order(X13[,1]),]
X15[order(X15[,1]),]
X16[order(X16[,1]),]

knitr::kable(round(cbind(X11[order(X11[,1]), c(1, 3:6)], 
                         X13[order(X13[,1]), c(1, 3:6)], 
                         X15[order(X15[,1]), c(1, 3:6)]), digits = 2), 
             col.names = 
               rep(c("Trt label","$x_{1}$", "$x_{2}$", "$x_{3}$", 
                     "$x_{4}$"), 3),
             escape = F, booktabs = T,
             linesep = "",
             caption = "Compound optimal design for $k = 4$ two-level 
                        factors with $n = 12$ runs and $\\kappa_{LP} = 
                        \\kappa_{LoF-LP} = \\kappa_{MSE(L)} = 1/3$, 
                        along with the corresponding $LP$- and 
                        $MSE(L)$-optimal designs.") |>
  kableExtra::add_header_above(header = c("Compound 1" = 5, 
                                          "$LP$-optimal" = 5, 
                                          "$MSE(L)$-optimal" = 5), 
                               escape = F)

# Figure 1 of criteria values for different designs and kappa
LP_plot <- pb_results |>
  subset(K < 11) |>
  ggplot(aes(x = K, y = LP, colour = design, shape = design)) +
  geom_jitter(height = 0, width = .1) +
  labs(x = "k")
MSE_plot <- pb_results |>
  subset(K < 11) |>
  ggplot(aes(x = K, y = MSE, colour = design, shape = design)) +
  geom_jitter(height = 0, width = .1) +
  labs(x = "k")
DF_plot <- pb_results |>
  subset(K < 11) |>
  ggplot(aes(x = K, y = DF, colour = design, shape = design)) +
  geom_jitter(height = 0, width = .1) +
  labs(x = "k", y = "PE")
Ls_plot <- pb_results |>
  subset(K < 11) |>
  ggplot(aes(x = K, y = L, colour = design, shape = design)) +
  geom_jitter(height = 0, width = .1) +
  labs(x = "k")
LP_plot + MSE_plot + DF_plot + Ls_plot + plot_layout(guides = "collect")
\end{CodeInput}

\end{document}